\documentclass[useAMS,usenatbib]{mn2e}

\usepackage{graphicx}
\usepackage{float}
\usepackage{pdfpages}
\usepackage{longtable}
\usepackage{lipsum}
\usepackage{textcomp,gensymb}
\usepackage{psfig}

\newcommand{\photu}{ph\ cm$^{-2}$ s$^{-1}$ sr$^{-1}$ \AA$^{-1}$}
\newcommand{\galex}{{\it GALEX}}

\title[Dust Scattering from the Taurus Molecular Cloud]{Dust Scattering from the Taurus Molecular Cloud}

\author[Sathyanarayan et al.]{Sathyanarayan$^{1}$\thanks{sathyanarayank@gmail.com}, Jayant Murthy$^{2}$\thanks{jmurthy@yahoo.com } and K. Narayanankutty$^{1}$ \\
$^{1}$Amrita Vishwa Vidyapeetham, Kollam 690525, India\\ 
$^{2}$Indian Institute of Astrophysics, Bengaluru 560034, India\\
}

\begin{document}
\date{Accepted . Received ; in original form }
\pagerange{\pageref{firstpage}--\pageref{lastpage}} \pubyear{2015}
\maketitle
\label{firstpage}

\begin{abstract}
We present an analysis of the diffuse ultraviolet (UV) emission near the Taurus Molecular Cloud based on observations made by the {\it Galaxy Evolution Explorer} (GALEX). We used a Monte Carlo dust scattering model to show that about half of the scattered flux originates in the molecular cloud with 25\% arising in the foreground and 25\% behind the cloud. The best-fit albedo of the dust grains is 0.3 but the geometry is such that we could not constrain the phase function asymmetry factor ($g$).
\end{abstract}

\begin{keywords}
Taurus Molecular Cloud, ultraviolet: ISM
\end{keywords}

\section{Introduction}

The Taurus Molecular Cloud (TMC) is part of the larger Taurus-Auriga complex of molecular clouds which extends over more than 100 square degrees in the sky. It is one of the nearest star forming regions at a distance of only 140 pc 
from the Sun with a thickness of around 20 pc \citep{Kenyon1994, Loinard2005, Lombardi2010}. Although it is near the Galactic Plane (b $\sim$ -14), there is little dust either in front of or behind the molecular cloud \citep{Padoan2002, Lombardi2010}. The TMC has been studied intensively at many different wavelengths and has provided a laboratory for the study of molecular cloud evolution and star formation in fine spatial and spectral detail. 

High resolution optical maps show that the TMC is comprised of loosely associated diffuse filaments mixed with highly clumpy molecular cores \citep{Panopoulou2014}. The clumpy molecular gas cores in TMC have a density of about few hundred $cm^{-3}$ \citep{Pineda2010} and produce very few low mass stars \citep{Whittet2004,Palmeirim2013} with a stellar density of about 1 - 10 stars pc$^{-3}$ \citep{Luhman2000}. In contrast to other star-forming regions such as Orion or Ophiuchus, the TMC is quiet with a low rate of star formation \citep{Kenyon1995} and no massive stars and hence no ionizing radiation \citep{Walter1991}.

The first diffuse UV studies of the TMC were made by \citet{Hurwitz1994} using the BEST (Berkeley Extreme Ultraviolet Shuttle Telescope) instrument, which flew on the Space Shuttle (STS-61C) in 1986 as part of the UVX payload. He observed both continuum emission due to starlight scattered by interstellar dust and emission lines from H${_2}$ fluorescence. The continuum emission was  inversely correlated with the optical depth suggesting that the diffuse background originated behind  and was shadowed by the dense molecular cloud; a conclusion later supported by observations from the SPEAR/FIMS instrument \citep{Lee2006}.

We have used observations made by the Galaxy Evolution Explorer (GALEX) in two bands (far-ultraviolet (FUV: 1350 - 1750 \AA) and near-ultraviolet (NUV: 1750 - 2850 \AA)) to study the diffuse UV radiation in and around the TMC. 
We will describe the observations and our modelling below. \citet{Lim2013} performed a similar analysis for a much larger region (the entire Taurus-Perseus-Auriga complex of molecular clouds). However, that included areas with quite different physical properties and we have chosen to focus on the TMC.

\section{Data}

\begin{figure*}
\includegraphics[width=7.4in]{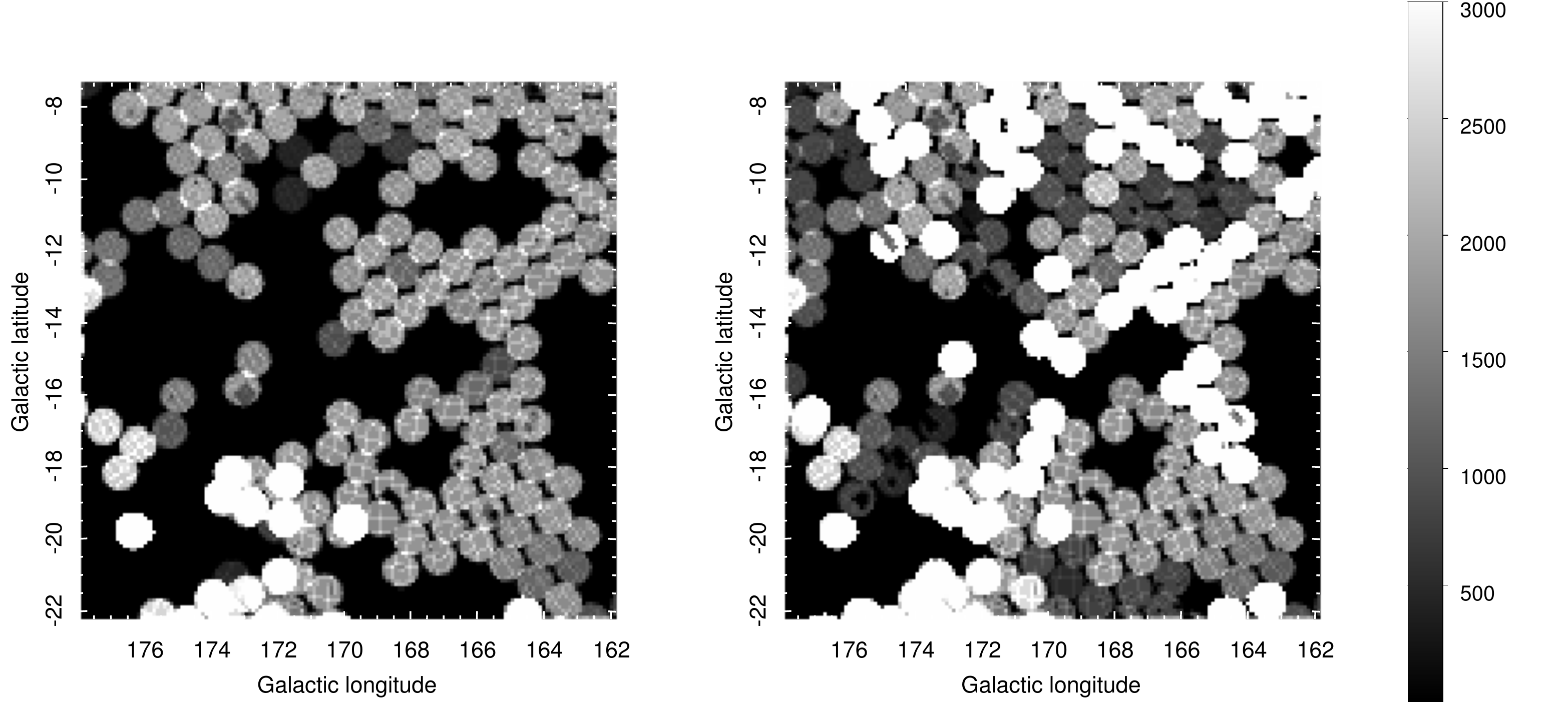}
\caption{\galex\ FUV and NUV total exposure time per pixel are shown from left to right. The exposure time are expressed in units of seconds.}
\label{fig_exposure}
\end{figure*}

\begin{figure*}
\includegraphics[width=2.3in]{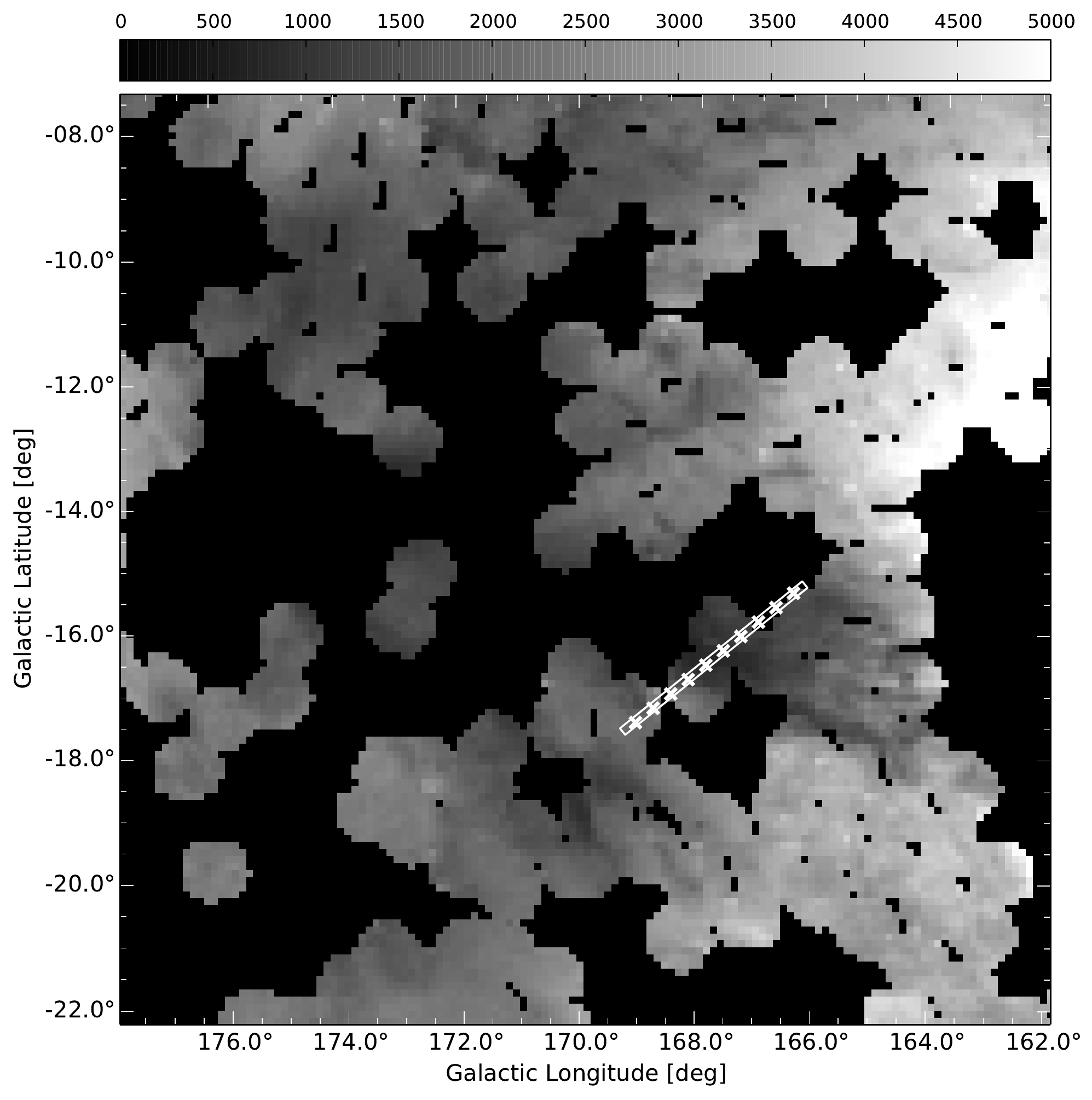}
\includegraphics[width=2.3in]{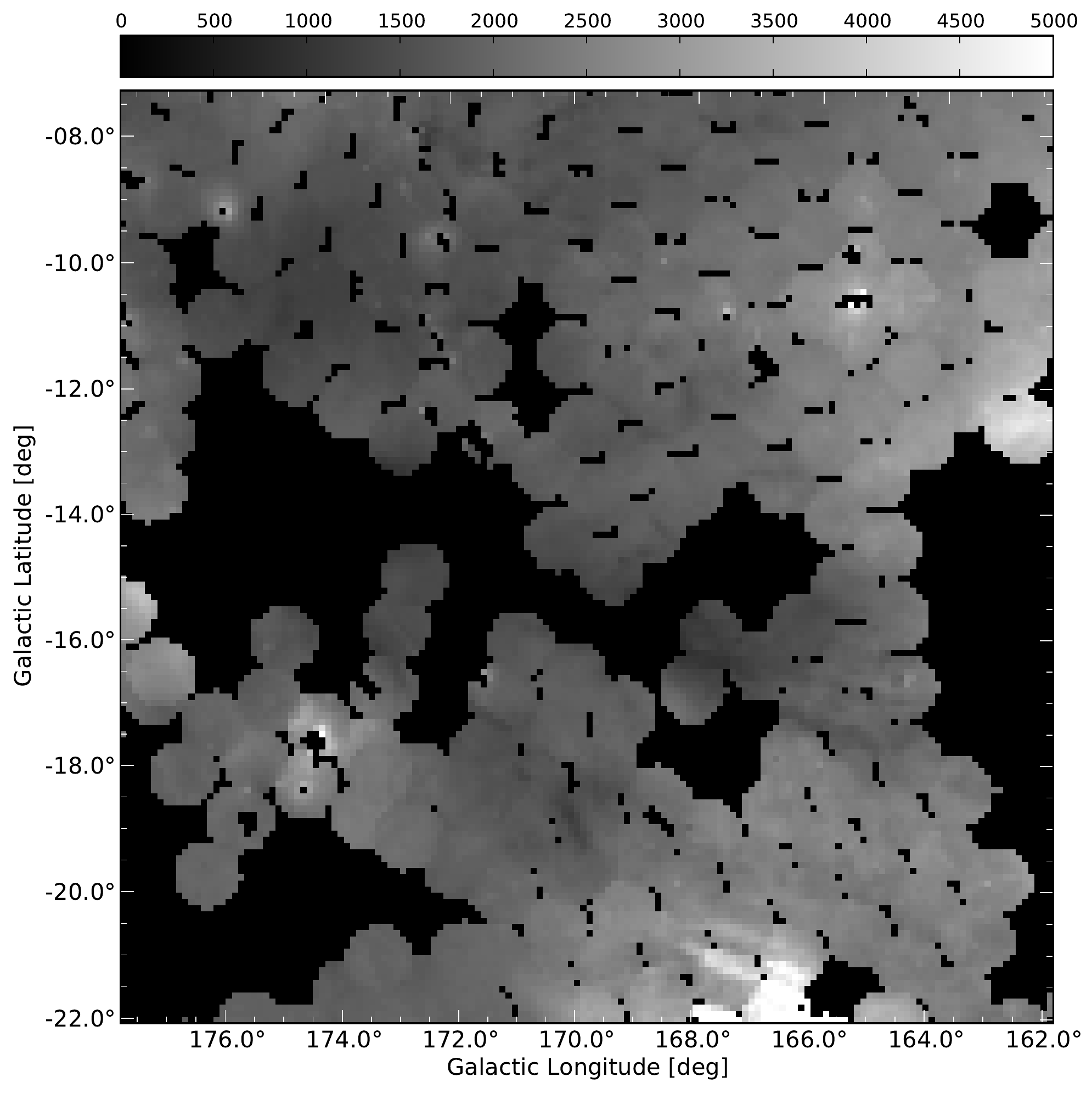}
\includegraphics[width=2.3in]{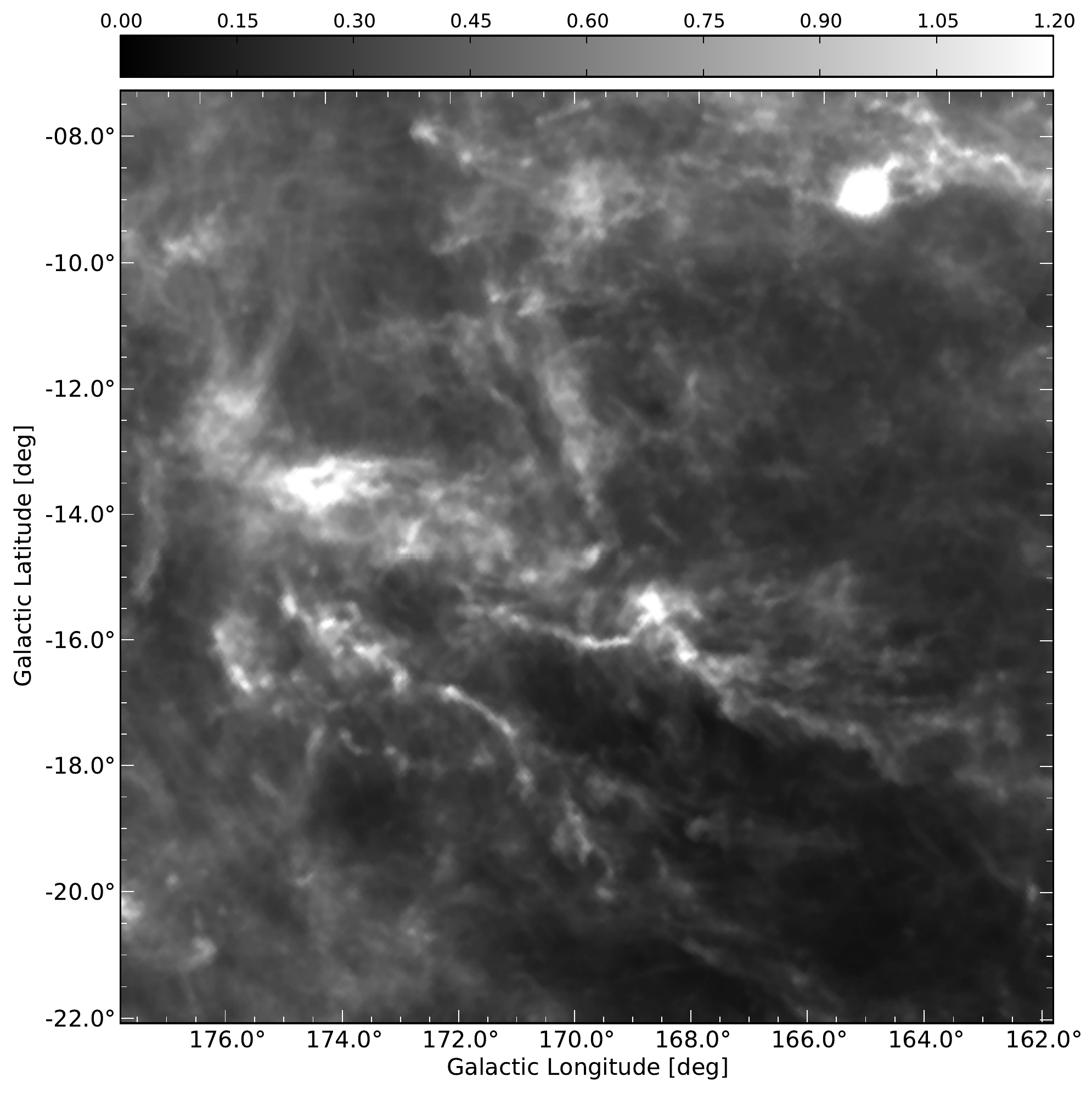}
\caption{\galex\ FUV and NUV and {\it Planck} E(B-V) images are shown from left to right with colour scales at the top. The UV images are in units of \photu\ and the E(B-V) is in magnitudes. The white crosses in the FUV image show the positions of the UVX observations \citep{Hurwitz1994}. Regions with no \galex\ observations appear as black in the UV images.}
\label{figimages}
\end{figure*}

GALEX, a NASA Small Explorer mission, was launched on April 28, 2003 with the primary goal of studying galaxy evolution at low red-shift \citep{Martin2005} and completed its mission on June 28, 2013 \citep{Bianchi2014}. The final data set of GALEX includes more than 44,000 observations \citep{Murthy2014b} in two ultraviolet bands. The data products from each observation have been described by \citet{Morrissey2007} and are archived at the Mikulski Archive for Space Telescopes (MAST).

We have shown an exposure time map of the observed region in the FUV and the NUV bands in Fig.~\ref{fig_exposure}. There were a total of 159 FUV and 258 NUV visits near the TMC region taken over a range of 9 years from 2003 to 2012 (2003 -- 2007 for the FUV). Most of the observations were taken as part of the AIS (All-sky Imaging Survey) with an exposure time of about 100 seconds per visit but with a few observations for specific science programs \citep[eg][]{Findeisen2010} with correspondingly greater exposure times. \citet{Murthy2014b} has calculated the diffuse Galactic background (DGL) values over the entire sky by masking out point sources and binning to a resolution of $2'$ and these data are available from the MAST archive\footnote{\it https://archive.stsci.edu/prepds/uv-bkgd/}.

We have downloaded all the data within 8 degrees of the nominal centre of the TMC (170.0, -14.0) and binned into 6$\arcmin$ bins, weighting each observations by its exposure time. Edge effects contaminate the outer part of the \galex\ detector and we have only used the central $0.5\degree$ (radius) of each observation. The resultant maps of the TMC in the FUV and the NUV are shown in Fig.~\ref{figimages} along with the extinction map derived from {\it Planck} observations \citep{planck2014}. There were no observations over significant parts of the TMC, partly because of safety concerns for the instrument \citep{Findeisen2010} and partly because of the failure of the FUV power supply.

\begin{figure}
\includegraphics[width=2.45in, angle=90]{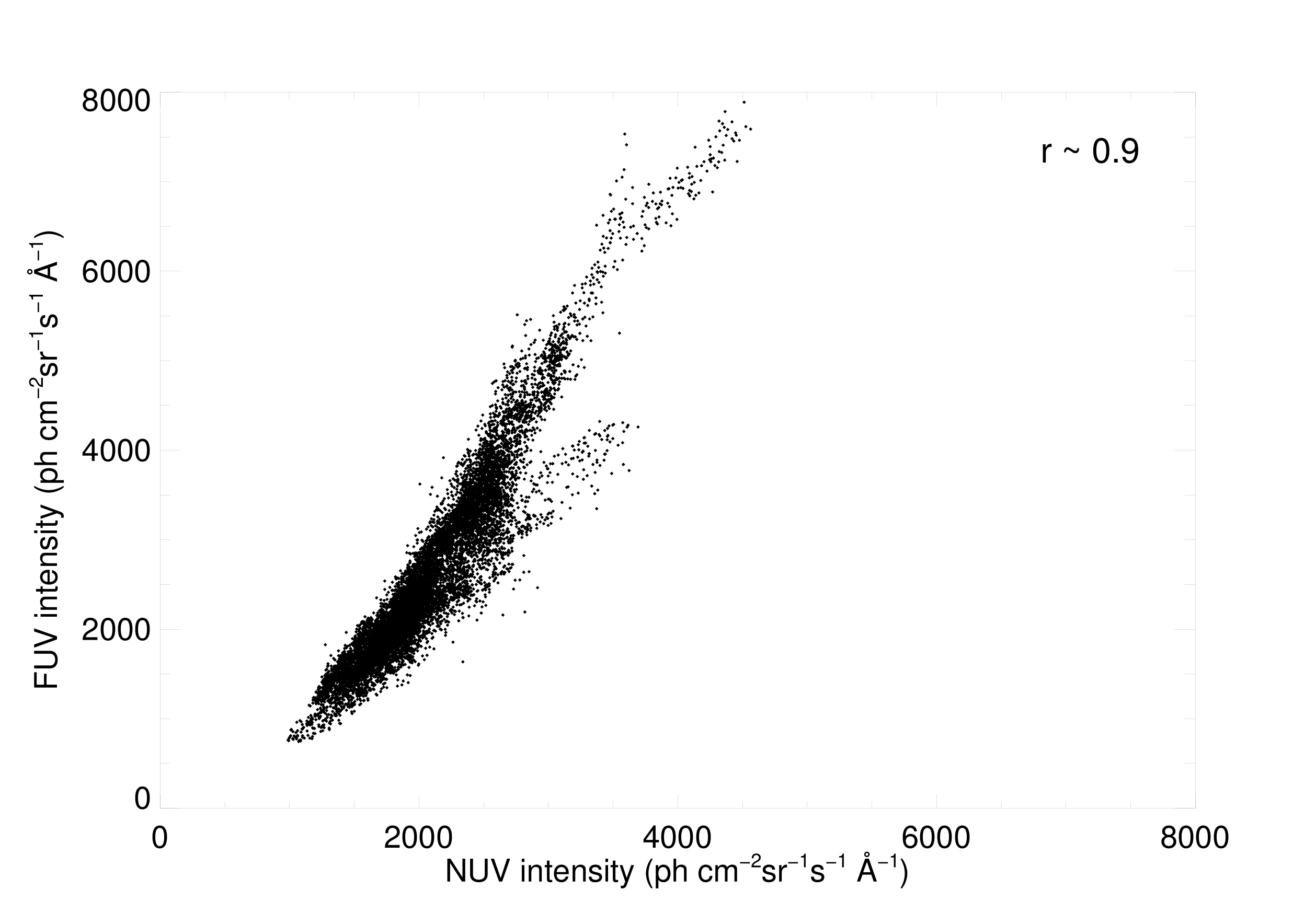}
\caption{Correlation between \galex\ FUV and NUV intensity.}
\label{fuv_nuv}
\end{figure}

\begin{figure}
\includegraphics[width=3.5in]{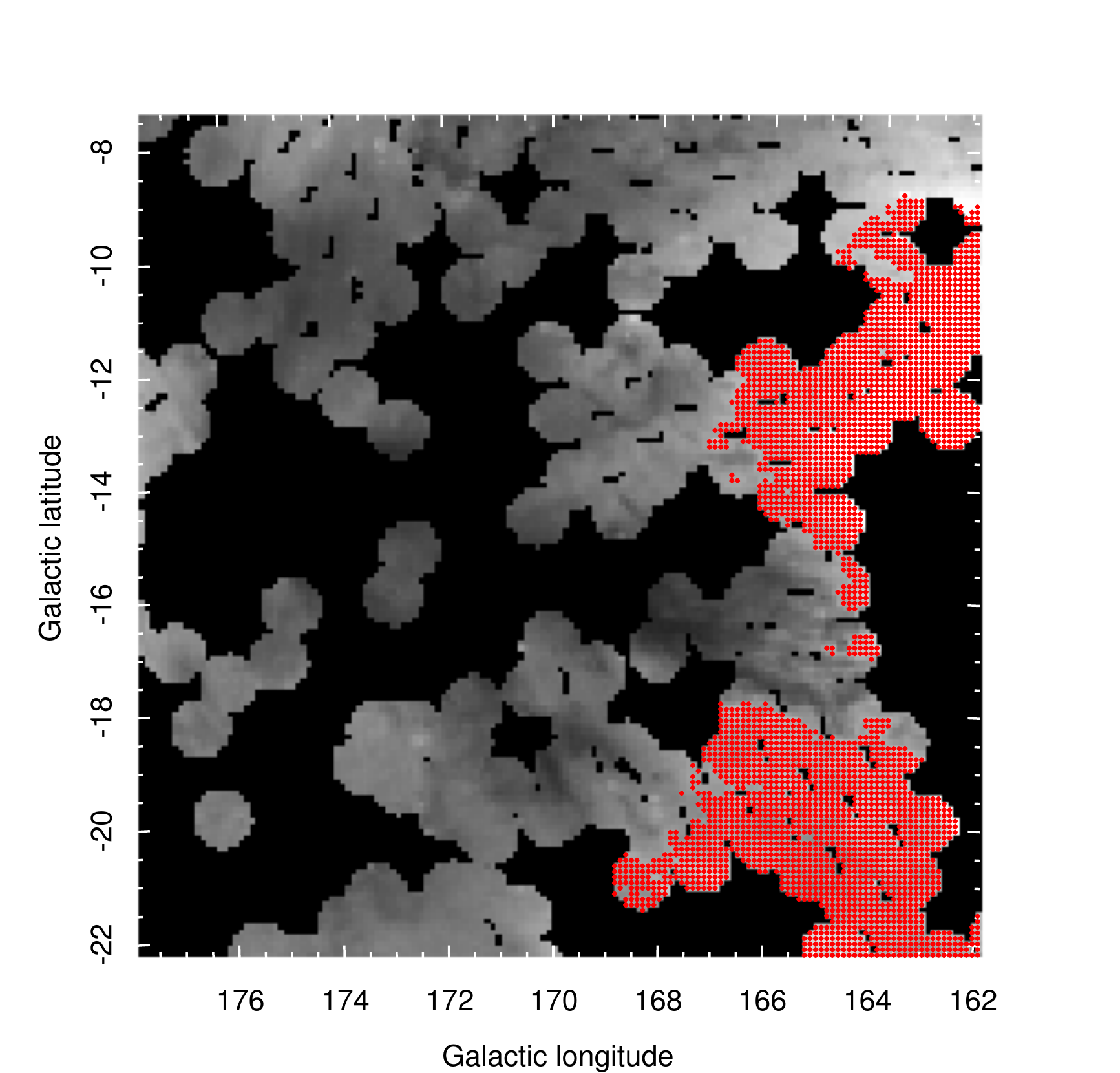}
\caption{Regions with different relationship between FUV and E(B-V). The dotted region in red colour on the right hand side of the image, is brighter in FUV and have little dust.}
\label{region1}
\end{figure}

\begin{figure}
\centering
\includegraphics[width=2.2in,angle=90]{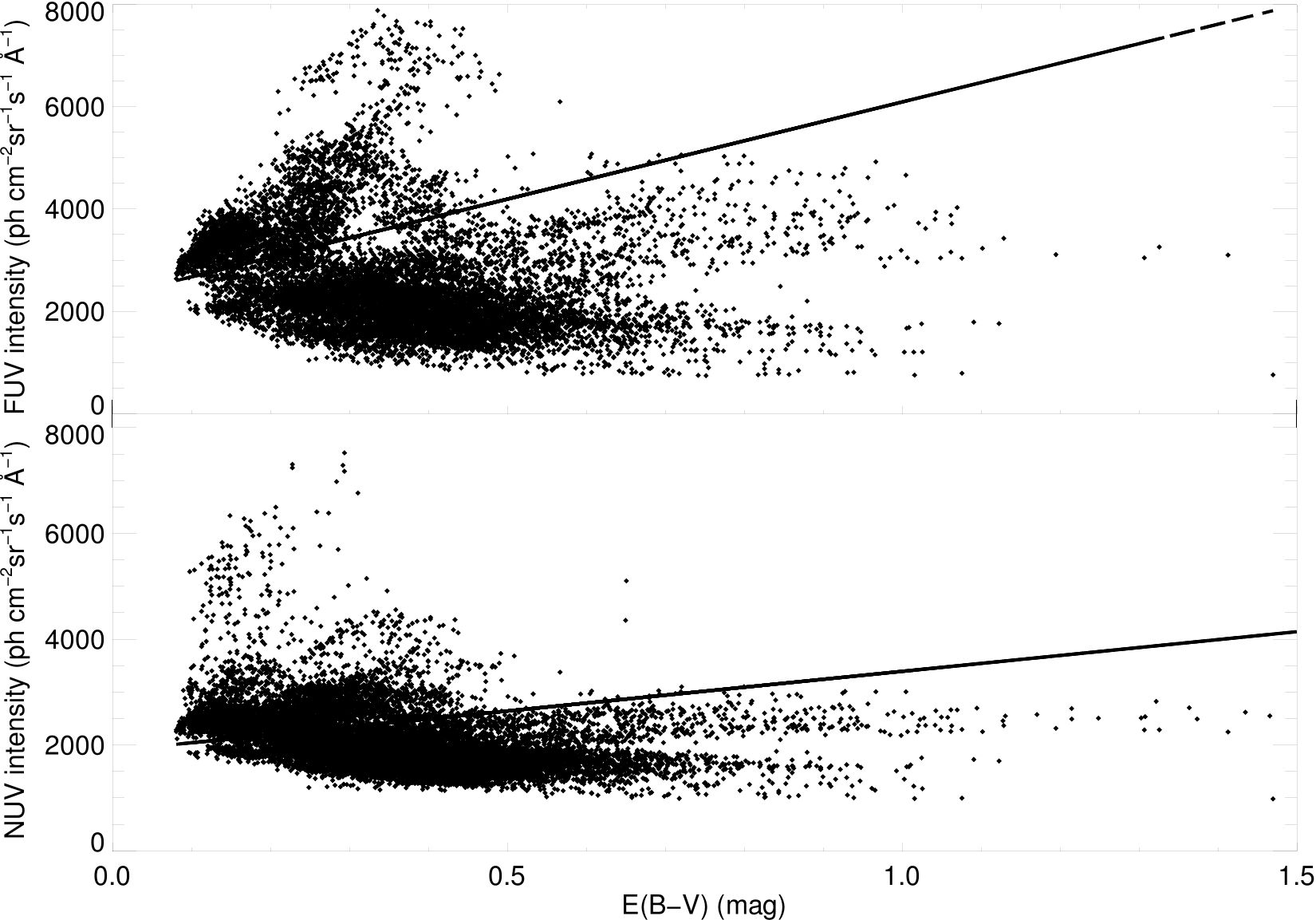}
\caption{\galex\ FUV (top) and NUV (bottom) intensity plotted against {\it Planck} E(B-V).The black line separates the regions having different relationship between UV and E(B-V).}
\label{uv_ebv}
\end{figure}

There is a strong correlation (r $\sim$ 0.9) between the FUV and the NUV (Fig.~\ref{fuv_nuv}). However, it is readily apparent that there is no correlation between the extinction and the diffuse UV background; rather, the UV images appear to be brighter to the right of the field where the extinction is least. This is borne out by a weak anti-correlation between the E(B-V) and the FUV (r = -0.25) and the E(B-V) and the NUV (r = -0.24). A more detailed examination of the data suggests that there exist two separate regions, demarcated in Fig. \ref{region1} where the points in red are those that fall above the line in Fig.~\ref{uv_ebv}. There is a positive correlation between the E(B - V) and the UV (FUV: r = 0.69; NUV: r = 0.51) in this region. The remaining region, which encompasses most of the TMC, has a negative correlation of r=-0.36 between the E(B - V) and the FUV and  r=-0.46 between the NUV and the E(B-V), as might be expected if the radiation were shadowed by the dust in the TMC.

The first observations of the diffuse UV in the TMC  was done by \citet{Hurwitz1994} using the UVX instrument on the Space Shuttle followed by observations made by using the SPEAR/FIMS \citep{Lee2006, Lim2013}. They observed the same anti-correlation between the reddening and the UV flux and interpreted this as due to the shadowing of a distant diffuse background by the molecular cloud. We will explore this in the next section.

\section{Modelling}

\citet{Murthy2016} has implemented a Monte Carlo model to predict the level of dust scattered radiation throughout the Galaxy. He used stars from the \textit{Hipparcos} catalog \citep{Perryman1997} along with model spectra for each spectral type \citep{Castelli2004} as sources for the stellar photons. Although we are interested only in the emission from the vicinity of the TMC, 
we have found that many of the scattered photons are from multiply scattered photons from outside the TMC and have therefore modelled the entire Galaxy, but with more detail in the vicinity of the TMC.

We have modelled the Galaxy as a 500x500x500 grid with a bin size of 2 pc on a side and filled with a hydrogen density of 1 cm$^{-3}$ at the Galactic Plane falling off with a scale height of 125 pc \citep{Marshall2006} from the plane. We calculated an optical depth for each bin using the cross-section per hydrogen atom tabulated by \citet{Draine2003} for ``Milky Way'' dust. 
\citet{Welsh2010} have found that there is a cavity of radius $\sim$ 80 pc (the Local Bubble) around the Sun and we have incorporated this in the dust model. However, this does not account for the complex structure in the TMC and we have separately modelled the dust in that region.

\begin{table*}
\centering
\resizebox{1.08\textwidth}{!}{\begin{minipage}{\textwidth}
 \caption{Properties of brightest stars near the TMC region with stellar flux contribution at three {\it g} values.}
\label{hip_stars_flux}

\begin{tabular}{cccccccccc}
  \hline
   HIP number & l & b & Spectral & V mag\footnote{From {\it SIMBAD}.} & Distance\footnote{From {\it Hipparcos catalogue}.} & \multicolumn{3}{c}{\% Total Flux \footnote{At constant {\it a} =0.3 and at 1500 \AA} } \\
 & ($\degree$) & ($\degree$) & type & &(pc) &  &  & \\
 & &  & & & &{\it g}=0.0&{\it g}=0.5&{\it g}=0.7 \\
 \hline
 
18246  &162.3 &-16.7 & B1&  2.84 & 301 &7.7 &14.7  & 23.8\\ 
18532  & 157.4&-10.1 &B1.5 &2.89 & 165& 7.7 &10.2 & 7.2\\
26727 & 206.5 &-16.6 &O9.7 &1.79 &226 &  4.3 & 4.3 & 2.2 \\
24661 & 166.2 & 2.2 &---&11.6 & 211& 3.1&4.1  &0.0 \\
26311 &205.2&-17.2 &B0 & 1.69& 412&  2.7  & 3.8 & 2.3 \\
25930  &203.9&-17.7&B0 &2.41 & 281& 2.3 & 2.8 & 1.5 \\
27366  &214.5&-18.5& B0.5& 2.06& 221&  2.3  &1.7 & 0.8 \\
4427 & 123.6 &-02.2 &B0.5 & 2.39&188 & 2.3 &1.2 &0.5\\
21881  & 176.7&-15.1 & B3& 4.26& 123& 2.2  &1.8 & 2.0\\
24436  & 209.2&-25.3 &B8 &0.13  & 237&  1.6  &1.3 & 0.9\\
17448 &160.4 &-17.7 &B1 &  3.84 & 452& 1.3 & 2.6 & 4.6 \\
17702  & 166.7& -23.5&B7 & 2.85& 113 & 1.3  & 0.9& 0.8\\
25336  & 196.9&-16.0 &B2 &1.64 & 75&  1.3 &0.3 &0.2\\
26451  & 185.7&-5.6&B1 &3.03 & 136&  1.2 &0.6 &0.5\\
60718  & 300.1&-0.4 &B1 &0.81 & 98& 1.2 & 0.3& 0.1\\
18614 & 160.4 &-13.1 &O7.5 & 4.01& 544&  1.1 & 3.3 & 4.0 \\
18724  & 178.4&-29.4 &B3 & 3.41& 114& 1.1  &0.5 & 0.0\\
30324  & 226.1&-14.3 &B1 & 1.97& 153& 1.0 &0.4 & 0.2\\
68702  & 311.8&1.3 &B1 &0.60 & 161& 1.0 &0.3 & 0.1\\
26207  & 195.1&-12.0 &O8 & 3.66& 324&  0.8 &1.4 & 0.9\\

\hline

\end{tabular}
\end{minipage}}
\end{table*} 

\begin{figure}
\centering
\includegraphics[width=2.5in,angle=90]{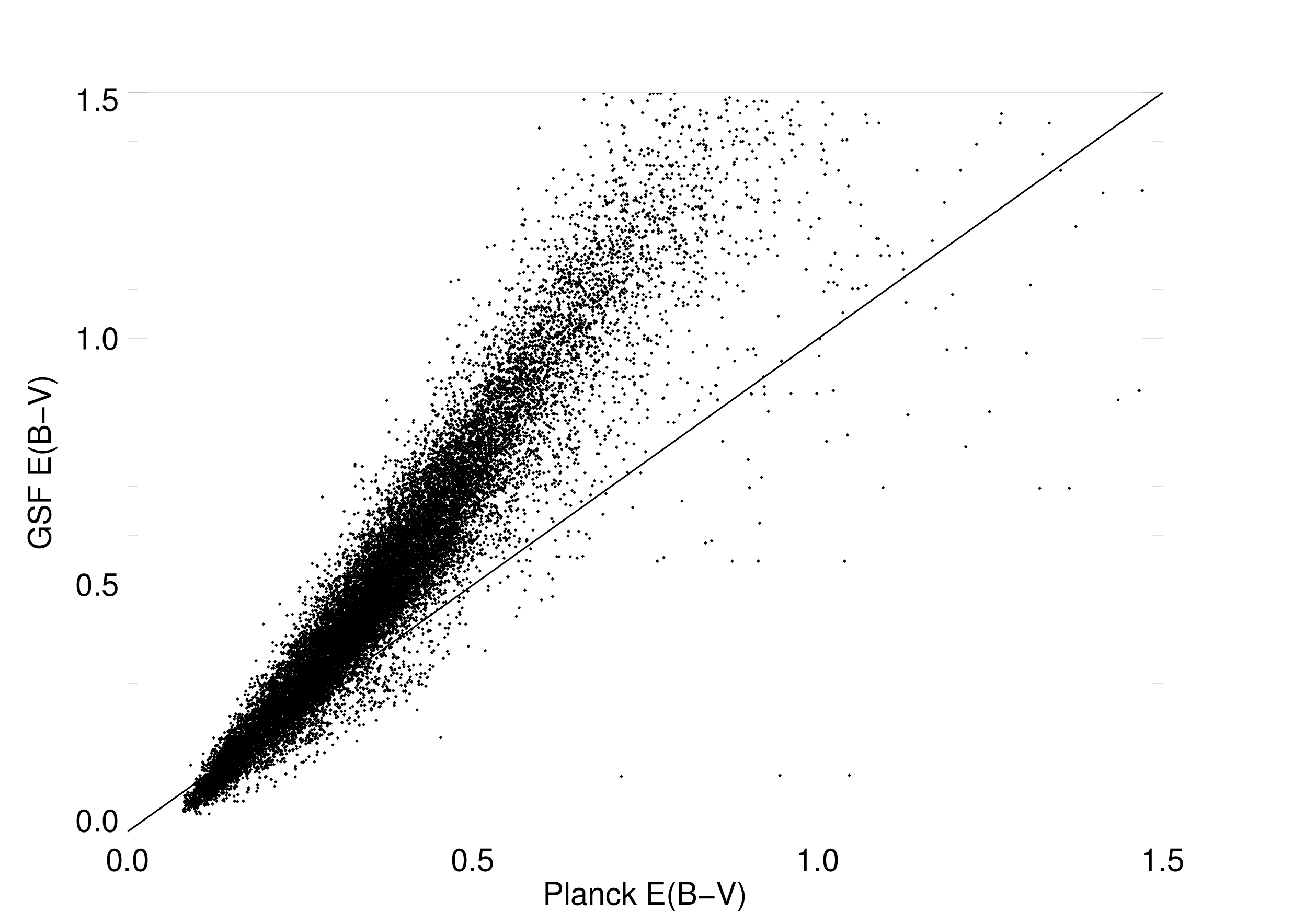}
\caption{Correlation between {\it Planck} and GSF reddening.}
\label{obs_green_planck}
\end{figure}
\begin{figure*}
\includegraphics[width=3.9in]{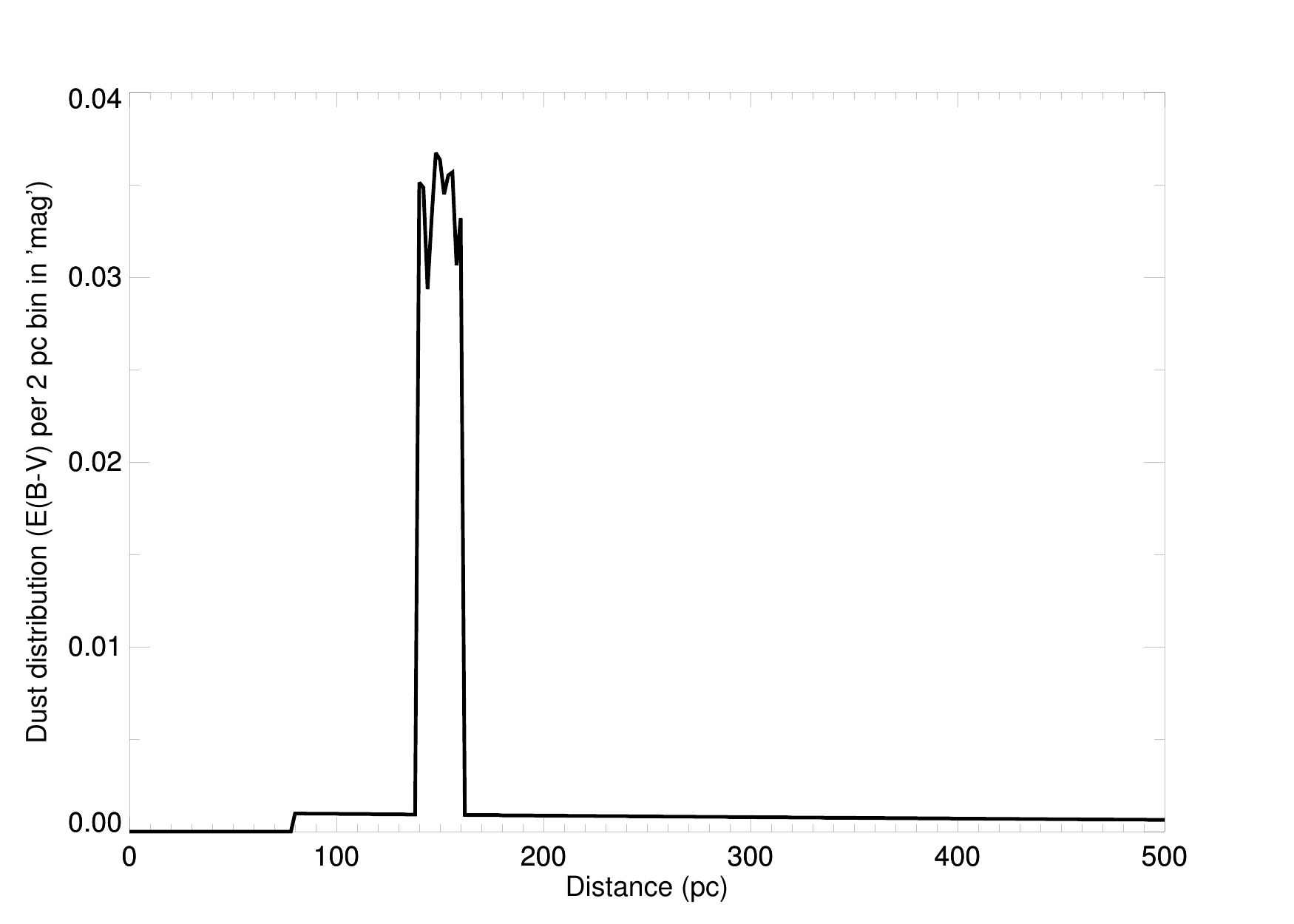}
\includegraphics[width=2.8in]{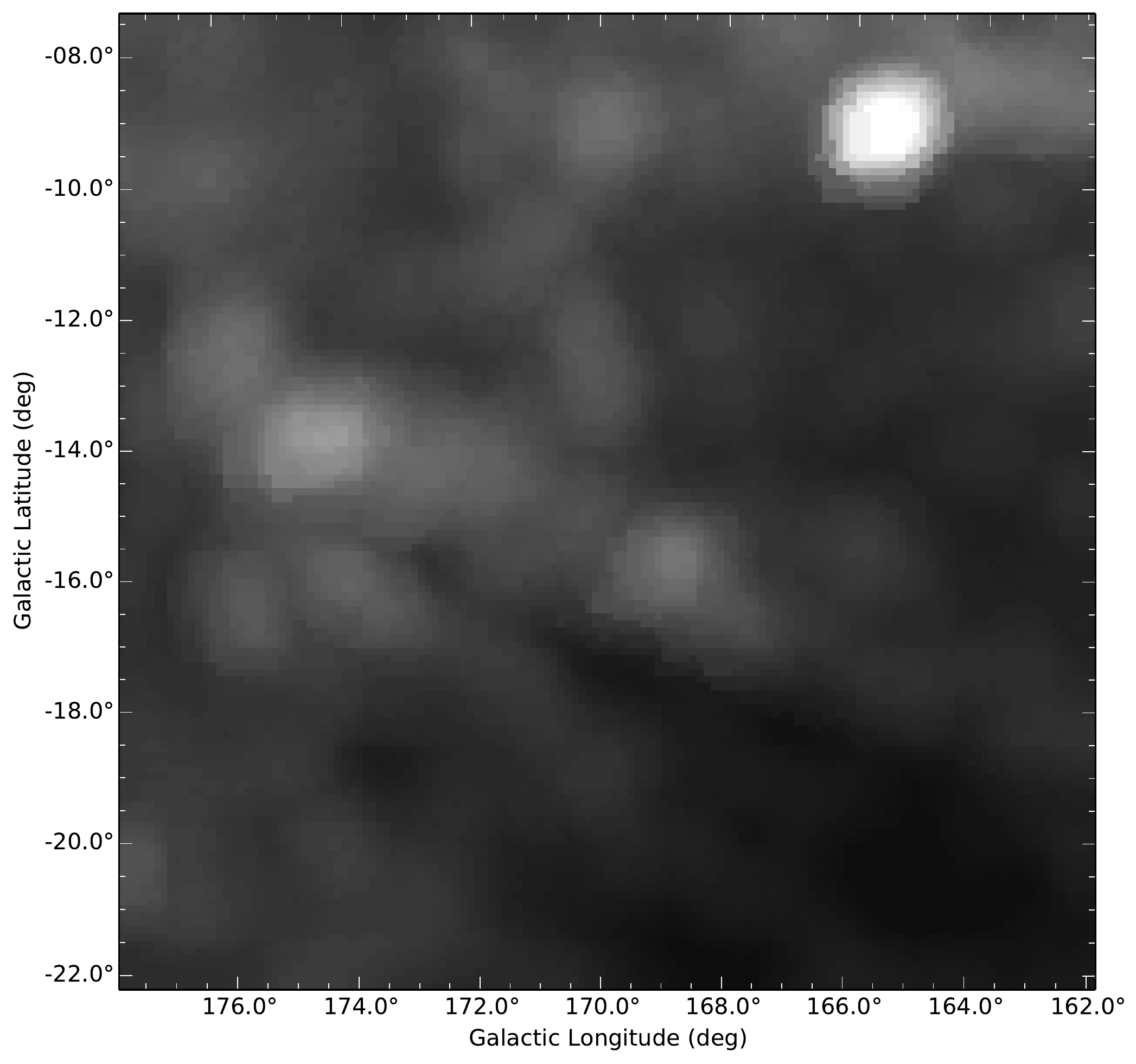}
\caption{Left: {\it Planck} dust distribution in terms of E(B-V) per 2 pc bin in units of `mag', as a function of distance along a representative line of sight. Right: Modelled spatial distribution of the dust in the region.}
\label{dust_plot_planck}
\end{figure*}

We first used the 3-dimensional extinction map of \citet{Green2015} (GSF hereafter) to characterize the dust distribution in the region finding that the bulk of the dust is found between 140 and 200 pc in reasonable agreement with the 140 -- 160 pc from earlier results \citep{Kenyon1994, Torres2007, Lombardi2010}. However, the total column density is greater than that derived from the {\it Planck} data \citep{planck2014} (Fig. \ref{obs_green_planck}). The GSF measurements are along specific lines of sight while the {\it Planck} measurements average over the $5'$ beam of the instrument and the larger values may be an indicator of clumping in the TMC. We have therefore used a second model where we have taken the reddening along any line of sight from the {\it Planck} data \citep{planck2014} and distributed the dust uniformly between 140 - 160 pc (Fig. \ref{dust_plot_planck}). Although this gives somewhat lower column densities than the GSF data, it does match the extinction measurements of \citet{AG_b1999}. We tested the effects of both models on our derived diffuse flux finding little difference between the two. We therefore continue with only using the dust distribution as given by the {\it Planck} model.

As has been standard in studies of interstellar dust, we used the Henyey-Greenstein scattering function \citep{Henyey1941} 
with the albedo ({\it a}) and the phase function asymmetry factor({\it g}) as free parameters to determine each scattering. 

\section{Results}
\begin{figure}
\includegraphics[width=2.5in, angle=90]{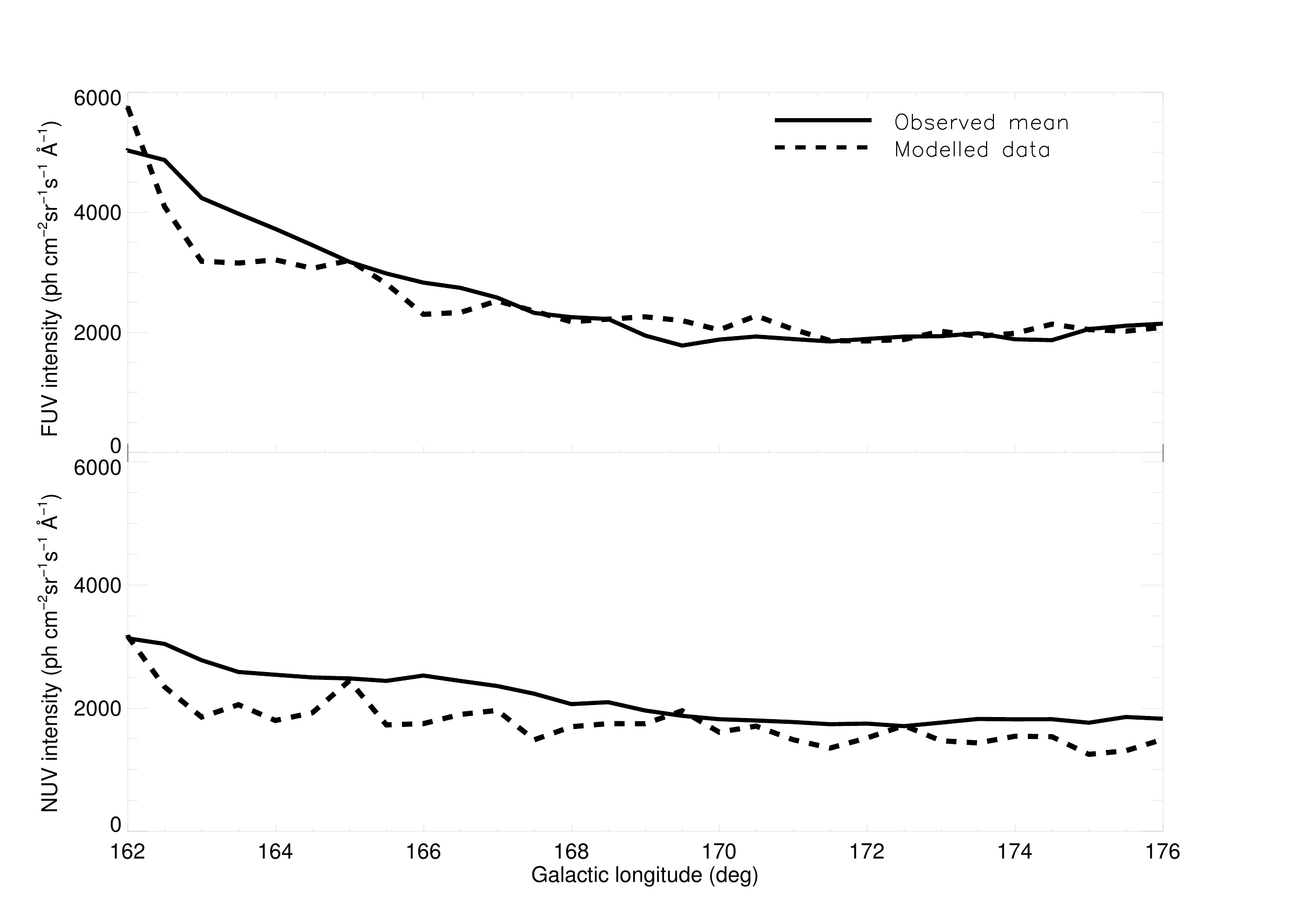}
\caption{Modelled FUV and NUV emission as a function of galactic longitude for $a$ = 0.3 and $g$ = 0.7, over-plotted with the original FUV and NUV emission.}
\label{fit_best}
\end{figure}

\begin{figure}
\includegraphics[width=3.6in]{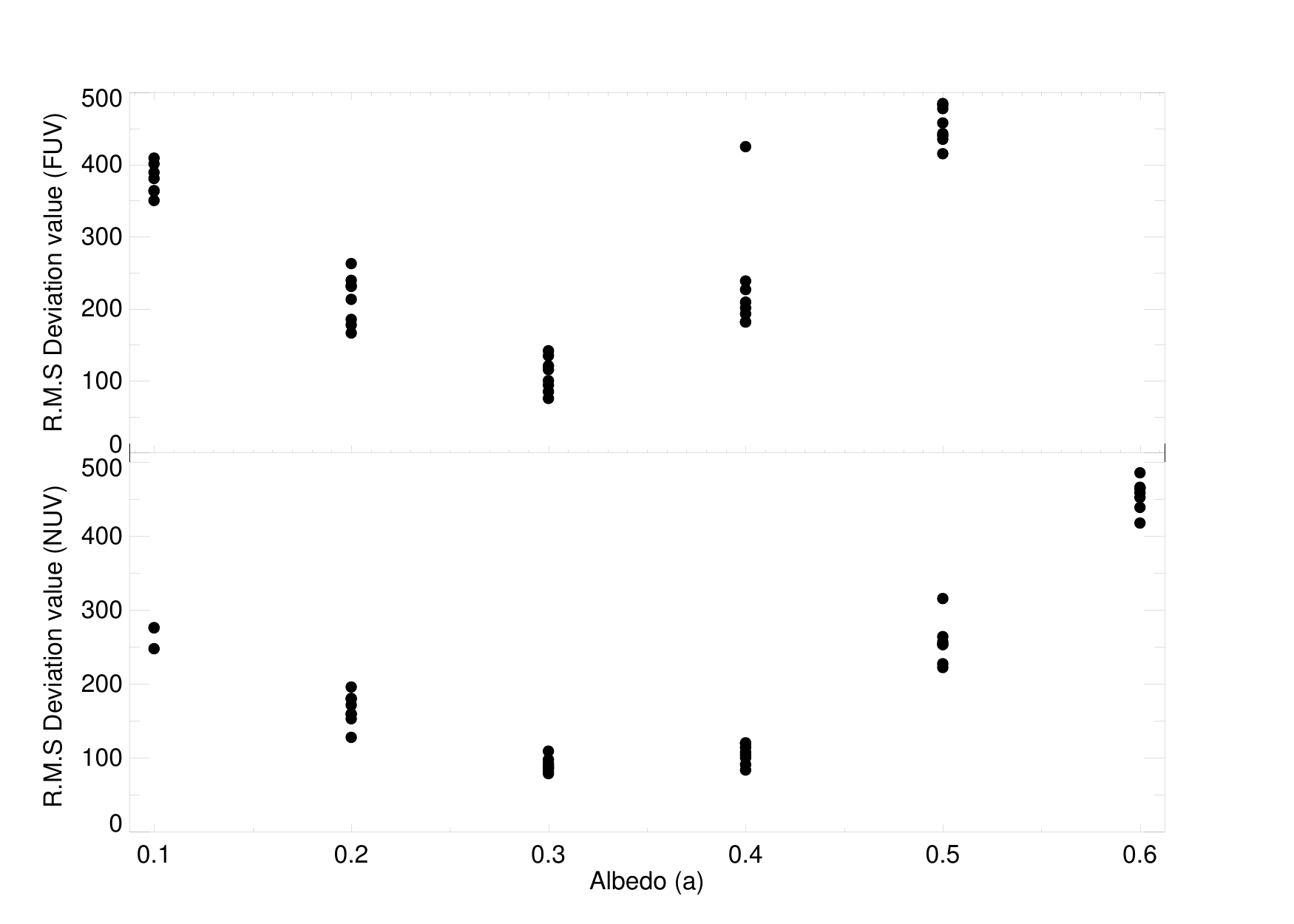}
\includegraphics[width=3.6in]{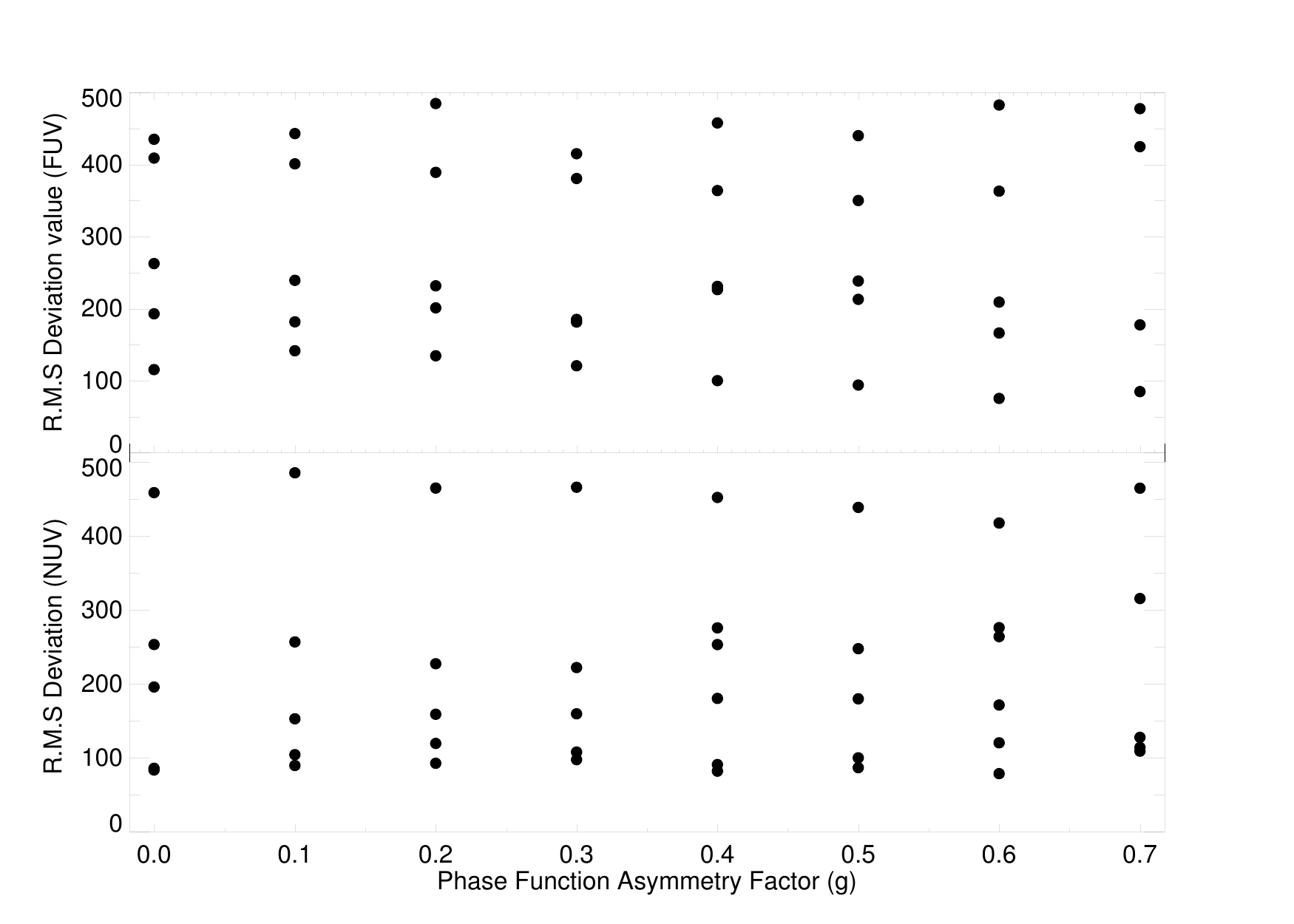}
\caption{Root mean square deviations as a function of albedo ({\it a}) and as a function of phase function asymmetry factor ({\it g}).}
\label{chisq_planck}
\end{figure}

We have run a series of Monte Carlo models to predict the diffuse background in both the FUV and the NUV bands for different combinations of {\it a} and {\it g}. We used a baseline of 100 million ($10^{8}$) photons for the entire grid but used simulations with 1 billion ($10^{9}$) photons for values of $a$ and $g$ near the best fit values. We cannot compare the models directly to the \galex\ data on a pixel by pixel basis because of the noise intrinsic to the Monte Carlo procedure and therefore used the root-mean-square (R.M.S) deviation of the flux as a function of Galactic longitude as a metric of the goodness of fit of the model (Fig.~\ref{fit_best}). We have plotted the  R.M.S. deviations for both the FUV and the NUV as a function of $a$ and $g$ in Fig.~\ref{chisq_planck} finding that the best fit occurs for an albedo of about 0.3 in both bands. Although this is in agreement with \citet{Hurwitz1994} and \citet{Lee2006} in the same region, it is lower than the albedo found in other parts of the Galaxy (compiled in Table 4 of \citet{Murthy2016}). \citet{Witt1996} showed that a clumpy medium, as in the TMC \citep{Panopoulou2014}, gives rise to a lower effective albedo and it is likely that the ``true'' albedo is greater than our derived value.

We have listed the 20 stars which contribute the most to the diffuse flux at $g = 0$ in Table \ref{hip_stars_flux}. The two stars which contribute the most flux at any value of $g$ are near the right edge of Fig. \ref{figimages} accounting for part of the rise in the UV fluxes to lower longitudes. However, there are enough stars contributing with a range of distances and positions that the dependence on $g$ is washed out, as was found by  \citet{Gordon1994} in the upper Scorpius region. We have adopted $g = 0.7$ which is consistent with most observations in the UV \citep{Draine2003}.

\begin{figure}
\includegraphics[width=3.6in]{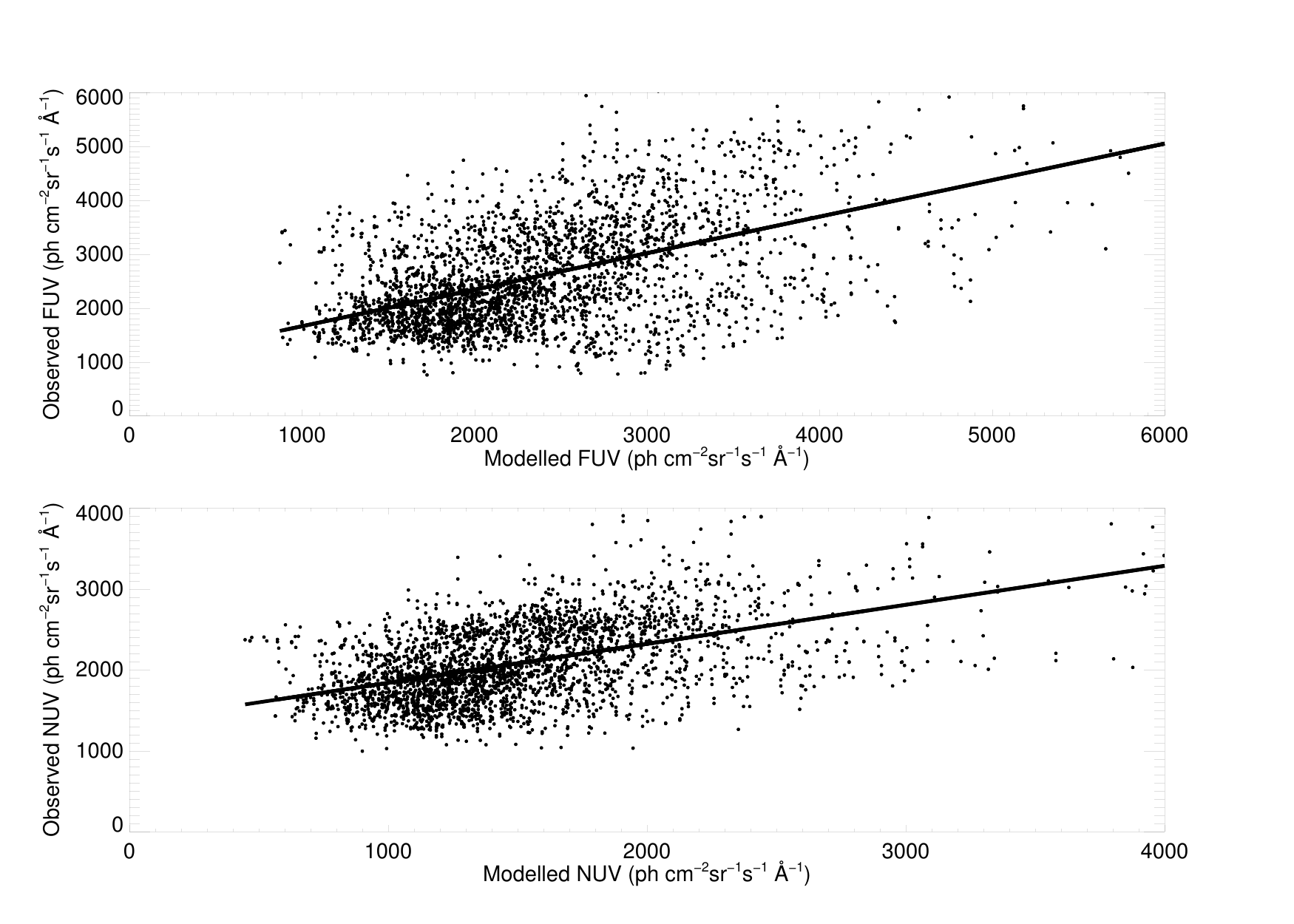}
\caption{Correlation plot between the modelled and the observed flux. The best fit lines are shown.}
\label{region1_2}
\end{figure}

\begin{figure}
\includegraphics[width=3.4in]{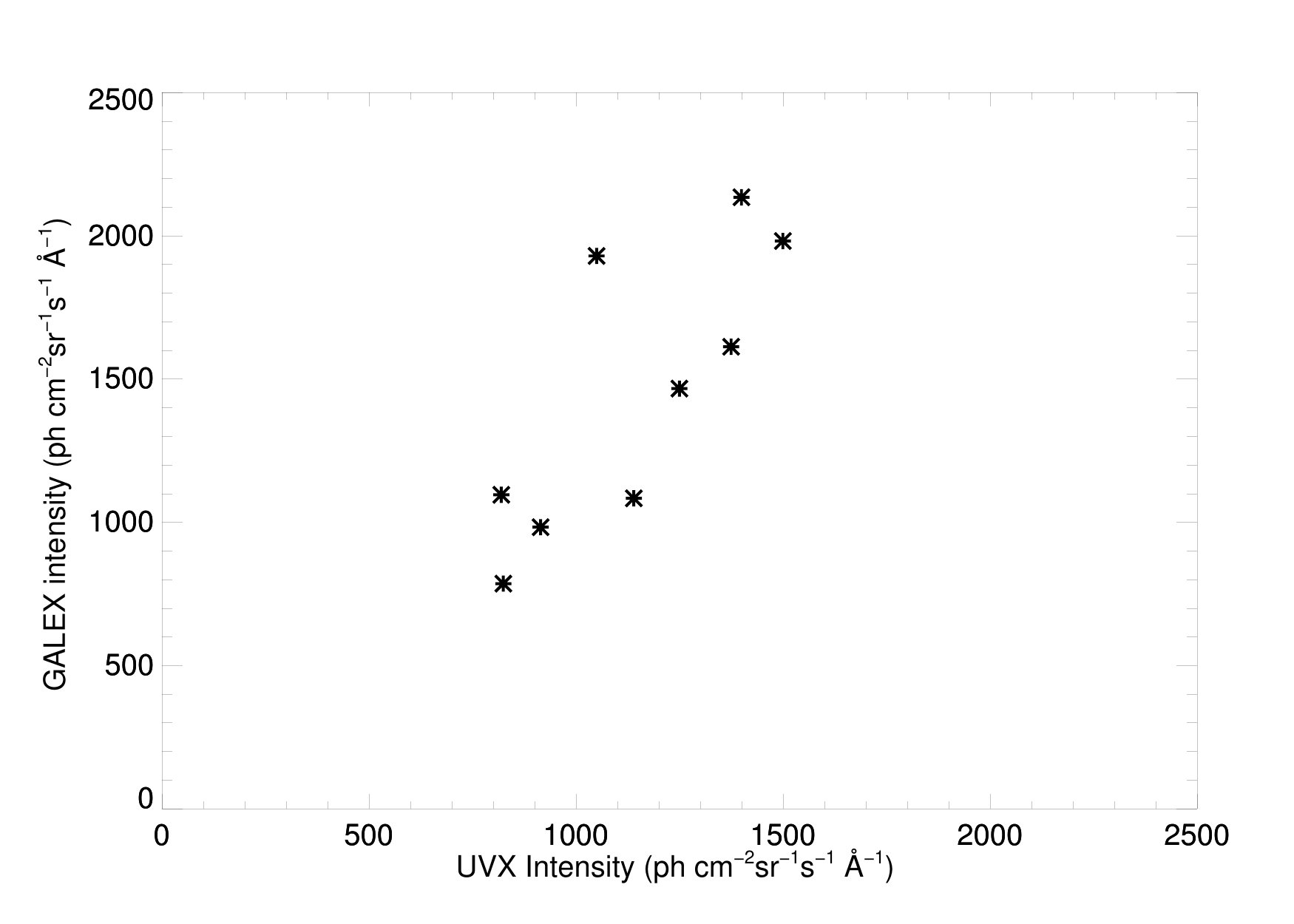}
\caption{Comparison of absolute values between the UVX values from \citet{Hurwitz1994} and the \galex\ intensity. 300 \photu\ have been subtracted from the UVX data to account for molecular hydrogen emission.}
\label{galex_hurwitz}
\end{figure}

\begin{figure}
\includegraphics[width=3.3in]{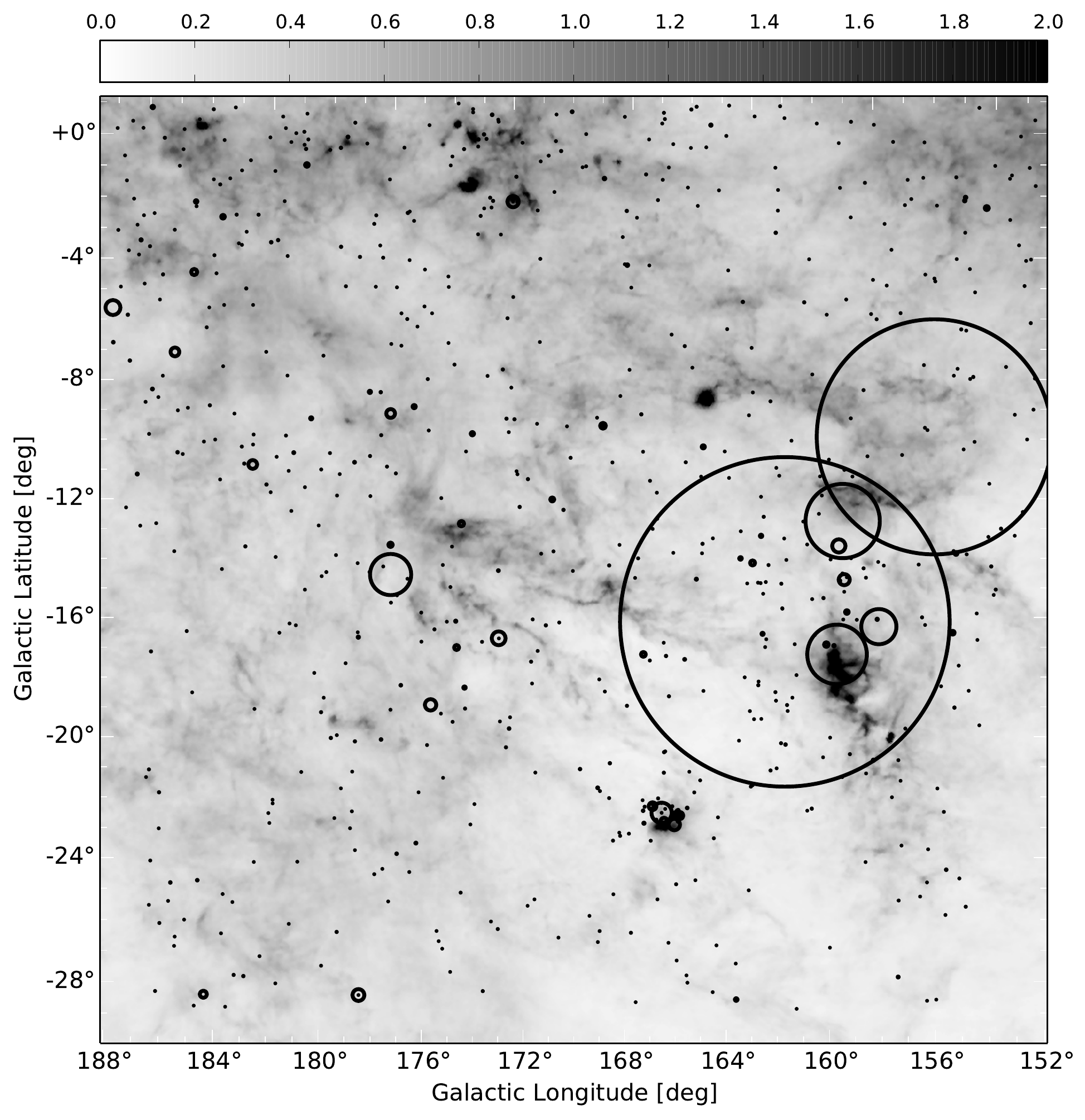}
\caption{Bright stars are marked as circles over  the Planck dust  reddening map. The centre of the circles represents the location of the stars while its radius proportional to contribution towards total flux.}
\label{circles_stars}
\end{figure}

\begin{figure}
\includegraphics[width=3.6in]{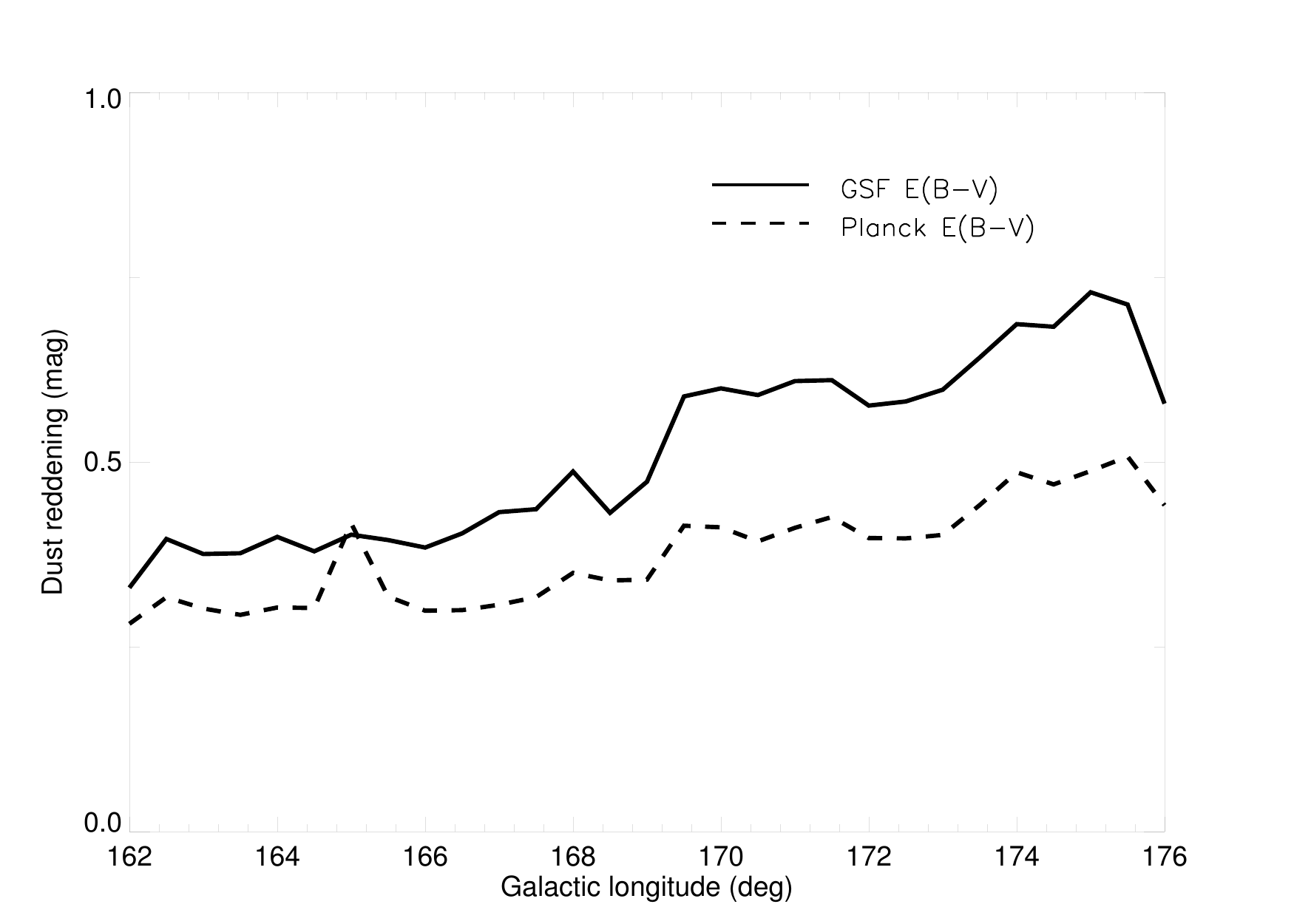}
\caption{Mean E(B-V) as function of galactic longitude in both GSF and {\it{Planck}} cases.}
\label{ebv_gl}
\end{figure}

\begin{figure}
\includegraphics[width=3.5in]{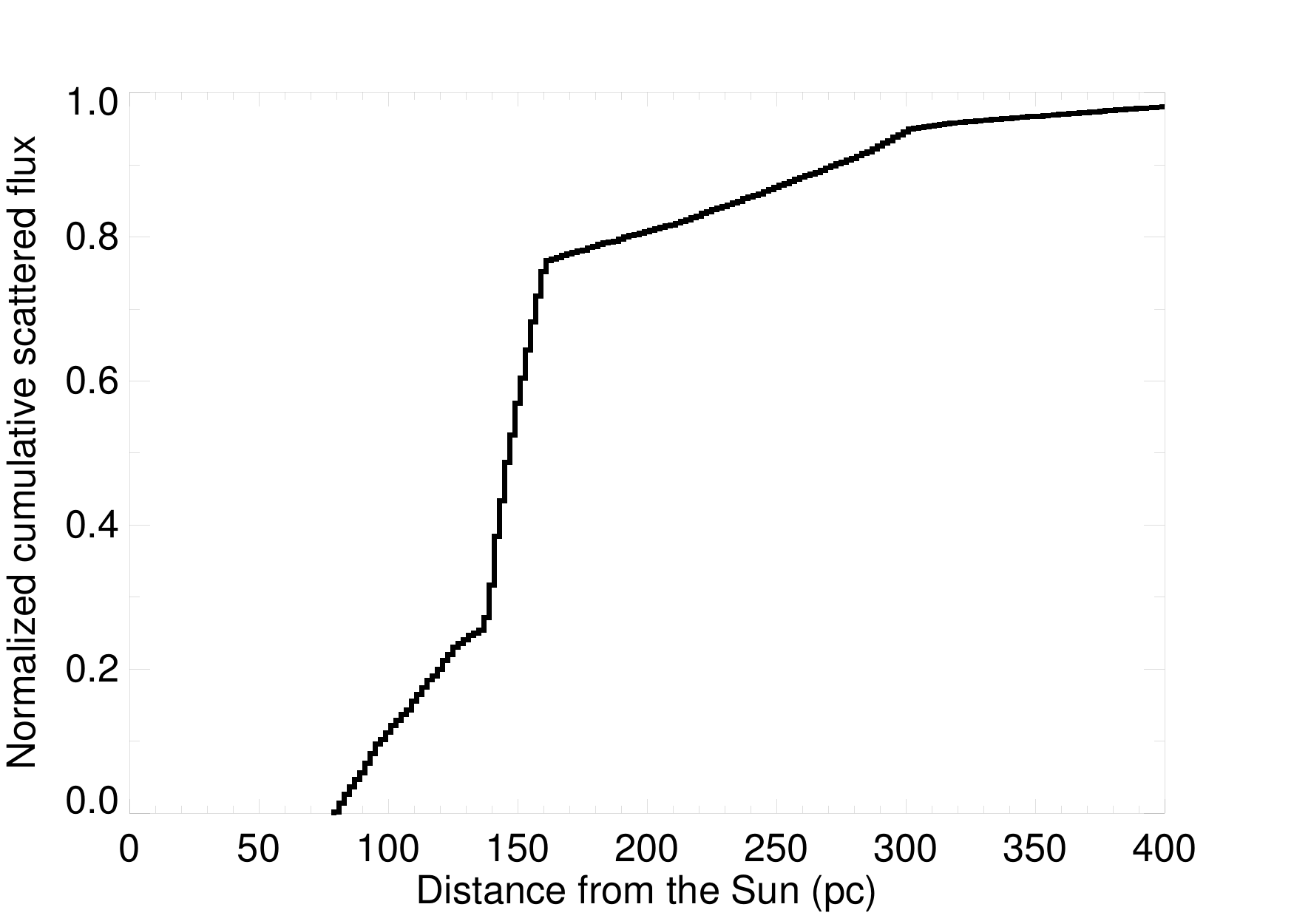}
\caption{Fraction of total FUV scattered flux in the TMC as a function of distance from the Sun.}
\label{cumflux_dist}
\end{figure}

We have run our model with $10^{10}$ photons for the best fit optical constants  ($a = 0.3; g = 0.7$) to improve the signal-to-noise and plotted the modelled versus the observed fluxes in Fig. \ref{region1_2}, where we have binned the data into $0.2^{\circ}$ bins. The correlation between the observed and modelled FUV is 0.55 with a slope of 0.68 and an offset of 980 \photu. The slope for the NUV is 0.48 with an offset of 1360 \photu\ and a correlation coefficient of 0.51. If we look at these values in the broader context of the diffuse radiation at low latitudes, \citet{Murthy2016} found that there was considerable scatter between the observed reddening and the UV flux, which he attributed to an uncertain dust distribution. We believe that the main source of scatter between our modelled fluxes and the data is the complex structure of the TMC. However, we do find offsets similar to the 700 --- 900 \photu\ found by \cite{Murthy2016} including 200 --- 300 \photu\ from airglow \citet{Murthy2014a}, molecular hydrogen fluorescence in the FUV \citep{Hurwitz1994,Lim2013} and a possibly new, unknown component \citep{Henry2015, Hamden2013}.

Finally, we return to the anti-correlation between the diffuse flux and the reddening observed by \citet{Hurwitz1994} and \citet{Lee2006}, which they interpreted as scattering from a more distant diffuse source extincted by the TMC. Our data agree with both the UVX observations of \citet{Hurwitz1994} (Fig. \ref{galex_hurwitz}) and the SPEAR observations of \citet{Lee2006} but we believe that the actual situation is more complex. At least part of the brightening to lower longitudes is coincidental because the two strongest contributing stars (Table \ref{hip_stars_flux}) are near a longitude of 160$^{\circ}$ (Fig. \ref{circles_stars}) where the reddening is less (Fig. \ref{ebv_gl}). These two B stars (HIP 18246 and HIP 18532) contribute ~45\% of the flux at a Galactic longitude less than 165$\degree$. 

Our model predicts that about 25\% of the total scattered flux is from dust in front of the TMC and another 25\% behind the cloud with the remaining 50\% arising in the TMC itself (Fig. \ref{cumflux_dist}), as might be expected given that almost all of the dust is in the TMC. However, our model is not well-suited to explore the scattering within a dense molecular cloud where clumping may be important. HIP 18246 is behind the TMC, as well as the other bright stars which make much of the field unobservable by \galex\ \citep{Findeisen2010}, and much of the flux from within the molecular cloud is due to the scattering of the light from those stars by dust within the cloud. Self-shielding within the cloud, itself, will yield the observed anti-correlation, exacerbated by the coincidental position of the two bright stars at one edge of the field.

\section{Conclusions}

We have modelled the diffuse radiation in the direction of the TMC using {\galex} data with a Monte Carlo model finding an albedo of $\approx 0.3$, but with no constraints on $g$. The albedo is likely to be an underestimate because we do not take clumping into effect.

About half of the scattered radiation originates in the body of the cloud with self-shielding giving rise to an anti-correlation between the observed flux and the reddening. The two brightest stars in the field are located in the direction of least reddening which adds to the anti-correlation.

We find offsets of about 1000 \photu\ in both bands  of which some part may be due to foreground contributors such as airglow and some to an unknown component \citep{Henry2015}.
 
\section*{Acknowledgements}
This work is based on the data from NASA's GALEX spacecraft. GALEX is operated for NASA by the California Institute of Technology 
under NASA contract NAS5-98034. The dust extinction map obtained from \textit{PLANCK} (http://www.esa.int/Planck), 
an ESA science mission with instruments and contributions directly funded by ESA Member States, NASA, and Canada. We thank the anonymous referee for constructive suggestions which has resulted in a better paper. Sathyanarayan would 
like to thank Dr. Sujatha and Ms. Jyothi for their help during the early stages of this work.

\nocite{*}
\bibliographystyle{mn2e}
\bibliography{references}
\label{lastpage}
\end{document}